\pdfoutput=1
\documentclass{article}
\usepackage{spconfa4,amsmath,graphicx}
\usepackage{subcaption}
\usepackage{hyperref}
\usepackage{xspace}

\usepackage{amsmath}
\usepackage{amssymb}
\usepackage{bbm}
\usepackage{physics}
\usepackage{siunitx}
\usepackage{trfsigns}
\usepackage{pifont}
\usepackage{etoolbox}
\usepackage{multirow}
\usepackage{csquotes}
\usepackage[shortlabels]{enumitem}  %
\usepackage{cite}   %
\usepackage{hyperref}
\usepackage[capitalize]{cleveref}
\usepackage{xifthen}
\usepackage{balance} %
\usepackage{booktabs}
\usepackage{listings}

\usepackage{makecell}

\usepackage{url}
\usepackage[acronym]{glossaries}

\let\oldcite\cite
\renewcommand{\cite}[1]{%
\ifthenelse{\isempty{#1}}%
{\inred{[cite!]}}%
{\oldcite{#1}}%
}

\newcolumntype{H}{>{\setbox0=\hbox\bgroup}c<{\egroup}@{}}   %

\newcommand*{\thl}{\fontseries{b}\selectfont}

\robustify\thl
\robustify\fontseries

\newcommand{\inred}[1]{\textcolor{red}{#1}}

\glsdisablehyper
\newacronym{film}{FiLM}{Feature-wise Linear Modulation}
\newacronym{ipd}{IPD}{inter-channel phase difference}
\newacronym{tac}{TAC}{Transform, Average, and Concatenate}

\makeatletter
\renewcommand{\cdots}{%
  \hbox{%
    \baselineskip -2pt %
    \lineskiplimit\z@
    \kern 1 \p@ %
    \hbox{.}\hbox{.}\hbox{.}
  }%
}
\makeatother

\usepackage{tikz}
\usepackage{pgfplots}
\pgfplotsset{compat=1.16}
\usepgfplotslibrary{groupplots}
\usetikzlibrary{positioning,arrows,matrix,fit,calc,patterns,chains,scopes,shapes.multipart,decorations,decorations.markings,backgrounds,shadows.blur}

\definecolor{palette-1}{HTML}{1f77b4}
\definecolor{palette-2}{HTML}{ff7f0e}
\definecolor{palette-3}{HTML}{2ca02c}
\definecolor{palette-4}{HTML}{d62728}
\definecolor{palette-5}{HTML}{9467bd}
\definecolor{palette-6}{HTML}{8c564b}
\definecolor{palette-7}{HTML}{e377c2}
\definecolor{palette-8}{HTML}{7f7f7f}
\definecolor{palette-9}{HTML}{bcdb22}
\definecolor{palette-10}{HTML}{17becf}

\tikzset{
    line/.style={draw,black,thick,rounded corners=1mm,line cap=round},
    noshortarrow/.style={line,->},
    arrow/.style={noshortarrow,shorten >=.3mm},
    doublearrow/.style={arrow,<->, shorten <=.3mm},
    smallbox/.style={draw,black,thick,rounded corners=3,fill=white,align=center},
    box/.style={draw,black,thick,minimum height=3em,text depth=0.25ex,rounded corners=3,fill=white,align=center},
    nopadding/.style={minimum height=0,inner sep=1mm},
    signalbox/.style={draw,black,thin,rounded corners=1mm,minimum width=7mm, minimum height=4mm,inner sep=0},
    pbox/.style={box,fill=black!10},
    backgroundbox/.style={inner xsep=3mm, inner ysep=1mm, draw, dashed, rounded corners,fill=orange!10},
    branch/.style={inner sep=0.3mm,circle,fill=black},
    operator/.style={draw,circle,black,rounded corners,inner sep=0,fill=white},
    vertex/.style={draw,ultra thin,circle,black,rounded corners,inner sep=0.6mm,fill=gray,fill opacity=0.5},
    edge/.style={line,very thick,line cap=butt},
    pattern1/.style={pattern=north west lines,pattern color=palette-1},
    pattern2/.style={pattern=north east lines,pattern color=palette-2},
    pattern3/.style={pattern=crosshatch,pattern color=palette-3},
    buswidth/.style={path picture={\draw[black,-] (path picture bounding box.south west) -- (path picture bounding box.north east);}}
}

\pgfdeclarelayer{foreground}
\pgfdeclarelayer{background}
\pgfsetlayers{background,main,foreground}

\title{INVESTIGATING THE INTEGRATION OF SPATIAL INFORMATION IN FOUNDATION-MODEL-BASED SPEAKER DIARIZATION}
\name{Marc Deegen, Adrian Meise, Reinhold Haeb-Umbach\thanks{Computational resources were provided by NHR/PC2.}}
\address{Paderborn University, Communications Engineering Department, Germany\\
\small\ttfamily
    \{deegen, meise, haeb\}@nt.upb.de \\
    }

\begin{document}

\makeatother
\setlength{\abovedisplayskip}{5pt}
\setlength{\belowdisplayskip}{5pt}
\setlength{\textfloatsep}{7pt plus 2.0pt minus 0.0pt}
\setlength{\floatsep}{5pt plus 0.0pt minus 0.0pt}
\ninept

\maketitle
\begin{abstract}
Spatial information gleaned from multi-channel input has been shown to lead to improvements in meeting processing tasks like diarization and source separation. At the same time, diarization based on features extracted by large pretrained single-channel foundation models, such as WavLM, achieved state-of-the-art performance. 
This work compares three approaches to integrate spatial features into foundation model-based diarization systems: the cascade of a beamformer and a single-channel foundation model, a multi-channel foundation model, and the conditioning of the downstream network on explicitly extracted spatial features. Results show that the beamformer front-end is even detrimental to diarization performance in regions of overlapped speech, while best performance is achieved with the conditioning, demonstrating that the incorporation of explicit spatial features is a competitive approach to foundation-model-supported diarization. This approach is further subjected to a detailed error analysis showing that the conditioning system removes errors to a good extent that would occur when either only spectral or only spatial features were used. 
\end{abstract}


\section{Introduction}
Speaker diarization is an important task in multi-talker conversational speech recognition systems delivering annotations about who spoke when, which are both helpful for subsequent speech processing tasks and informative as ancillary annotations for users.
End-to-end neural diarization with vector clustering (EEND-VC) has emerged as a powerful approach to speaker diarization \cite{kinoshita21_icassp, kinoshita21_interspeech}. It performs EEND locally on short segments and subsequently aligns the segment-level predictions by clustering. 
The frameworks PyAnnote~\cite{bredin23_pyannote} and DiariZen \cite{han2025leveraging} are based on this approach and are widely used in the speech community. 

With the advances of self-supervised learning (SSL), large pretrained foundation models (FMs) like WavLM~\cite{Chen2022_WavLM} have  made inroads also in diarization systems due to the strong and robust frame-wise representations they extract from the audio. An example is the use of WavLM in the  DiariZen framework \cite{han2025leveraging},  a  Conformer-based~\cite{gulati20_interspeech} architecture to infer the speaker activities in an EEND-VC fashion.

It comes to no surprise that most current speech foundation models are trained on single-channel data, simply because much more single-channel data is available than multi-channel data. For example, WavLM is trained on 94k hours of speech data, while UniX-Enc \cite{unixenc}, a native multi-channel FM, is trained on roughly 700h of speech, more than a factor of 100 less.

However, many speech processing tasks benefit from multi-channel input. It is therefore an open question how to best combine foundation models with multi-channel processing. This question is particularly interesting for speaker diarization, because here the benefit of multi-channel processing can even be two-fold: first, the spatial diversity of the target signal and distortions can be exploited to suppress unwanted signal components, e.g., by beamforming, and, second, the different positions of the speakers in space can be used to more reliably decide who is speaking. 

In this contribution, we compare three approaches to leverage both multi-channel input and FMs for diarization. The first is a simple cascade of a beamformer with a single-channel FM-based feature extractor, prior to diarization. We opted for BeamformIt, which is known for its simplicity and robustness \cite{beamformit}. WavLM features are then computed from the beamformer output. The second is a multi-channel FM. While there exist first native multi-channel FMs \cite{unixenc}, we opted for a lightweight extension of WavLM, which introduced learnable inter-channel communication modules into the WavLM architecture, as proposed in \cite{han2025spatial}. The third employs a dedicated auxiliary multi-channel network to extract spatial features that are merged with WavLM features using them as conditioning information in Feature-wise linear Modulation (FiLM) layers, as proposed in \cite{hscma}.
All approaches are in common that they are designed to be independent of the number of microphones and the array configuration. Furthermore, using the same FM, WavLM, as the starting point, fosters comparability. The choice of WavLM was guided by its robustness and excellent performance on conversational speech processing tasks \cite{cspb}.

All three are examples of ``early fusion'' approaches. We will not consider ``late fusion,'' where diarization is conducted separately on each channel, and the channel-wise diarizations are afterwards fused to a single final diarization result, such as is done in Dover-Lap \cite{Raj2021Doverlap}. Late fusion tends to be significantly more computationally demanding, while being often less effective, as was found in \cite{han2025spatial}. Furthermore, early fusion can be applied to speech processing tasks other than diarization, but that is beyond the scope of this paper. 

The aforementioned third approach, the explicit computation of spatial and single-channel spectral features, which are then combined, is further scrutinized for some explainability investigations. We carry out an error analysis into how spectral and spatial information contribute to improved diarization performance. Such statistics may pave the way for further improving system performance.

Related to our study is the conversational speech processing benchmark for self-supervised speech models \cite{cspb}. It compares several FMs w.r.t. their performance on automatic speech recognition, speaker diarization, and enhancement/separation on four conversational speech corpora. The authors found that BeamformIt $+$ WavLM Large achieved the best overall performance. While their focus was on comparing FMs, our emphasis is different: We adopt their best performing configuration as one approach (though in the base instead of large configuration of WavLM) and compare it with other combinations of multi-channel input and FMs for diarization, and find that their best configuration is not the best one here.

\section{Integration of multi-channel information}
\label{sec:integration}
Three approaches are considered 
how to integrate multi-channel data and foundation model features to leverage jointly spatial and spectral information for diarization. 
 
\subsection{Cascaded approach}
\begin{figure}
    \begin{subfigure}{\columnwidth}
        \centering
        \begin{tikzpicture}[
    wav/.style={
        minimum width=6.2ex,
        minimum height=4ex,
        path picture={
            \foreach \size [count=\i] in {0.1, 0.2, 0.4, 0.7, 1, 0.7, 0.55, 0.5, 0.4, 0.6, 0.9, 0.4, 0.6} {
                \def\x{\i*.45ex}
                \def\y{\size*1.8ex}
                \path[top color=black!40,bottom color=black,rounded corners=1] ($(path picture bounding box.west) + (\x-.14ex,-\y)$) rectangle ($(path picture bounding box.west) + (\x+.14ex,\y)$);
            }
        }
    }
]
    \node[box,minimum width=5em] (wavlm) {\footnotesize Single-channel\\\footnotesize Foundation\\\footnotesize Model};
    \node[draw,right=1.4em of wavlm, fill=palette-2!10, align=center] (film1) { \footnotesize Downstream\\ \footnotesize model};

    \node[box,minimum width=4em, draw=black,
fill=none, left=3em of wavlm] (encoder) {\footnotesize Beam-\\\footnotesize former};

    \draw[arrow] (encoder) -- node[pos=0.5, anchor=south] {\resizebox{2.2em}{!}{\begin{tikzpicture}
                \node[
        minimum width=6.2ex,
        minimum height=4ex,
        path picture={
            \foreach \size [count=\i] in {0.1, 0.2, 0.4, 0.7, 1, 0.7, 0.55, 0.5, 0.4, 0.6, 0.9, 0.4, 0.6} {
                \def\x{\i*.45ex}
                \def\y{\size*2.0ex}
                \path[top color=black!40,bottom color=black,rounded corners=1] ($(path picture bounding box.west) + (\x-.14ex,-\y)$) rectangle ($(path picture bounding box.west) + (\x+.14ex,\y)$);
            }
        }
]{};
            \end{tikzpicture}}}
 (wavlm);
    
    \draw[arrow] (wavlm) -- node[pos=0,anchor=south west] {} (film1); %

    \def\wavoffset{2.0ex}
    \node[wav,left=2.25em of encoder, anchor=south] (wav2) {};
    \node[wav,yshift=-\wavoffset,  semitransparent] (wav3) at (wav2) {};
    \node[wav,yshift=-\wavoffset, nearly transparent] (wav4) at (wav3) {};

    \draw[arrow] (wav2.east) -- (wav2-|encoder.west);
    \draw[arrow] (wav3.east) -- (wav3-|encoder.west);
    \draw[arrow] (wav4.east) -- (wav4-|encoder.west);

\end{tikzpicture}
        \caption{Cascaded approach using a beamforming preprocessing that reduces the multi-channel input to single-channel output.}
        \label{fig:hybrid}
    \end{subfigure}
    
    \begin{subfigure}{\columnwidth}
        \centering
        \begin{tikzpicture}[
    wav/.style={
        minimum width=6.2ex,
        minimum height=4ex,
        path picture={
            \foreach \size [count=\i] in {0.1, 0.2, 0.4, 0.7, 1, 0.7, 0.55, 0.5, 0.4, 0.6, 0.9, 0.4, 0.6} {
                \def\x{\i*.45ex}
                \def\y{\size*1.8ex}
                \path[top color=black!40,bottom color=black,rounded corners=1] ($(path picture bounding box.west) + (\x-.14ex,-\y)$) rectangle ($(path picture bounding box.west) + (\x+.14ex,\y)$);
            }
        }
    }
]

    \node[box,minimum width=7em, fill=white] (wavlm) {\footnotesize Multi-channel\\\footnotesize Foundation\\\footnotesize Model};

    \node[draw,right=1.75em of wavlm, fill=palette-2!10, align=center] (film1) {\footnotesize Downstream\\\footnotesize model};

    \draw[arrow] (wavlm) -- node[pos=0,anchor=south west] {} (film1); %

    \def\wavoffset{2.0ex}
    \node[wav,left=3em of wavlm, anchor=south] (wav2) {};
    \node[wav,yshift=-\wavoffset,  semitransparent] (wav3) at (wav2) {};
    \node[wav,yshift=-\wavoffset, nearly transparent] (wav4) at (wav3) {};

    \coordinate (wavcenter) at ($(wav2)!.5!(wav3)$);

    \draw[arrow] (wav2.east) -- (wav2-|wavlm.west);
    \draw[arrow] (wav3.east) -- (wav3-|wavlm.west);
    \draw[arrow] (wav4.east) -- (wav4-|wavlm.west);

\end{tikzpicture}
        \caption{Multi-channel foundation model that directly uses the multi-channel input to extract features used in the downstream model as in~\cite{han2025spatial}.}
        \label{fig:mc_ssl}
    \end{subfigure}   
    
    \begin{subfigure}{\columnwidth}
        \centering
        \begin{tikzpicture}[
    wav/.style={
        minimum width=6.2ex,
        minimum height=4ex,
        path picture={
            \foreach \size [count=\i] in {0.1, 0.2, 0.4, 0.7, 1, 0.7, 0.55, 0.5, 0.4, 0.6, 0.9, 0.4, 0.6} {
                \def\x{\i*.45ex}
                \def\y{\size*1.8ex}
                \path[top color=black!40,bottom color=black,rounded corners=1] ($(path picture bounding box.west) + (\x-.14ex,-\y)$) rectangle ($(path picture bounding box.west) + (\x+.14ex,\y)$);
            }
        }
    }
]
    \node[box,minimum width=5em] (wavlm) {\footnotesize Single-channel \\ \footnotesize Foundation\\ \footnotesize Model};
    \node[box,right=1.3em of wavlm] (film1) {\footnotesize FiLM};
    \node[box,right=0.8em of film1] (conf1) {\footnotesize Conf.\\ \footnotesize block};

    \node[right=0.6em of conf1, inner sep=0em, outer sep=0em] (dots) {$\cdots$};

    \node[draw,dashed,fit={(film1)(conf1)},inner sep=1.1ex, xshift=0.05em] (block) {};

    \node[anchor=north west,inner ysep = 0,inner xsep=1pt,] at (block.north east) {\footnotesize $\times N$};

    \draw[arrow] (wavlm) -- node[pos=0,anchor=south west] {} (film1); %
    \draw[arrow] (film1) -- (conf1);
    \draw[arrow] (conf1) -- (dots);

\node[box, draw=black, fill=none, dashed, left=1.75em of wavlm] (wav1) {\footnotesize Beam-\\ \footnotesize former};

\node[left=1.3em of wav1, wav] (single_channel) {};
\node[below=-0.05em of single_channel, align=center] {\footnotesize Ref. channel};

    \draw[arrow](single_channel) -- (wav1);
    \draw[arrow](wav1) -- (wavlm);
   
    \node[anchor=south west] at (wav1.north west) {};%

    \def\wavoffset{2.0ex}
    \node[wav,below=1.4em of wav1,  xshift=0em] (wav2) {};
    \node[wav,yshift=-\wavoffset,  semitransparent] (wav3) at (wav2) {};
    \node[wav,yshift=-\wavoffset, nearly transparent] (wav4) at (wav3) {};
    \node[anchor=south west] at (wav2.north west) {}; %
    \draw[arrow, dashed] (wav2) -- (wav1);

    \coordinate (wavcenter) at ($(wav2)!.5!(wav3)$);
    \node[box,minimum width=6em, draw=black,
fill=none] at (wavlm|-wavcenter) (encoder) {\footnotesize Auxiliary\\ \footnotesize Multi-channel\\ \footnotesize Network};
    \draw[arrow] (wav1.east) ++ (0.5em,0) node[branch]{} |- ($(encoder.west) + (0,1.5*\wavoffset)$);
    \draw[arrow] (wav2.east) -- (wav2-|encoder.west);
    \draw[arrow] (wav3.east) -- (wav3-|encoder.west);
    \draw[arrow] (wav4.east) -- (wav4-|encoder.west);

    \draw[arrow] (encoder) -| node[above,pos=0,anchor=south west] {} (film1); %

    \begin{pgfonlayer}{background}
    
        \node[draw,fit={(film1)(conf1)(dots)},fill=palette-2!10,inner sep=1.75ex,  inner ysep=2.2em, yshift=-1.05em] (conformer) {};
    \end{pgfonlayer}

    \node[anchor=south, below=-2.4em of conformer, align=center, xshift=1.15em] {\footnotesize Conditioned\\\footnotesize Downstream model};
\end{tikzpicture}
        \caption{Conditioning of the downstream model with spatial features extracted by an auxiliary multi-channel network as in~\cite{hscma}. Here, the downstream network uses Conformer architecture and the input to the foundation model is either the single-channel input or beamformer output.}
        \label{fig:hscma}
    \end{subfigure}
    \caption{Overview of different approaches to incorporate multi-channel information into foundation-model-based diarization.}
\end{figure}
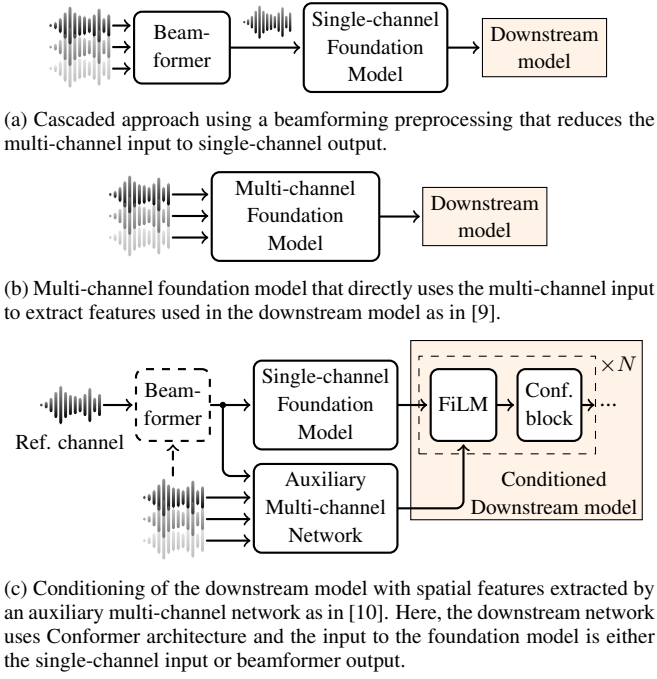

An obvious method to connect multi-microphone input to a single-channel FM is to prepend the single-channel FM by a beamformer, which merges the multiple channels to a single output stream, as shown in \Cref{fig:hybrid} and as proposed in \cite{cspb}.  Here, we employ BeamformIt \cite{beamformit}, a filter-and-sum beamformer based on Time Delay of Arrival estimates, that makes no assumptions about the number of channels and the array topology. 
No changes in the foundation model architecture are needed, %
which renders this approach a simple and straightforward way to accommodate multi-channel input.

\subsection{Multi-channel foundation model}
Here, the foundation model directly accepts the multiple microphone signals as input and extracts features from them, as shown in \Cref{fig:mc_ssl}. 
Models like UniX-Encoder \cite{unixenc} and the feature extractor of multi-channel DiariZen \cite{han2025spatial} are representatives of this approach. UniX-Enc is a dedicated model for multi-channel processing and mainly uses cross-channel transformers to capture information between the different channels. In our study, we consider the multi-channel extension of WavLM, specifically developed as a front end of the DiariZen diarization system to accommodate multi-channel input. It 
consists of several stacked WavLM layers, which are run in parallel on each input channel and which are connected via cross-channel communication blocks. 

The system can be pretrained on single-channel input, while the less abundant multi-channel data is then employed to train the channel communication blocks and fine-tune the whole system.
Model size and complexity of multi-channel FMs are typically increased compared to the cascaded approach.

\subsection{Multi-channel conditioning}
In \cite{hscma}, a third approach is proposed where first both multi-channel and FM-based single-channel features are extracted and then integrated, see \Cref{fig:hscma}. The FM is again the WavLM, and the spatial features are obtained by an auxiliary multi-channel network. It uses \glspl{ipd} of all non-redundant microphone pairs and the magnitude of the reference channel as input. The network consists of $5$  layers, where self-attention with shared weights followed by \gls{tac} \cite{luo2020end} connections are applied on each input in every layer. After averaging over all layers, the resulting spatial embedding is processed by a linear projection layer, layer normalization, and an additional conformer with classification head. 
This spatial network is pretrained on the downstream task, i.e., with a diarization objective, %
aiming to provide more discriminative and structured spatial cues, before its output is used as conditioning input to \gls{film} \cite{film} layers that combine the spatial and spectral features, as can be seen in \Cref{fig:hscma}. 
These \gls{film} layers are interspersed between the Conformer blocks of the DiariZen EEND module. Please refer to \cite{hscma} for details. 

Optionally, the reference channel, which is the input of the foundation model and whose magnitude is one input to the auxiliary multi-channel network, can be replaced by the output of a beamformer, thus possibly improving signal quality of this input. We will also investigate this option.

\section{Experiments}
\label{sec:experiments}

To have a reference if multi-channel information is beneficial at all, we adopt %
the single-channel DiariZen~\cite{han2025leveraging}. 
For the cascaded approach, the same setup is used except that the multi-channel input is first preprocessed with an additional beamformer. 
As example for the multi-channel FM, we use the multi-channel DiariZen system from~\cite{han2025spatial} for comparison, which is a multi-channel extension of the above baseline system. 
Results for the single- and multi-channel DiariZen are obtained by running the publicly available code\footnote{https://github.com/BUTSpeechFIT/DiariZen} in our environment.
The multi-channel conditioning system is adopted from \cite{hscma} and we additionally evaluate it with a beamforming step, whose output serves as an alternative to the single reference channel.

\begin{sloppypar}
All systems are trained on the multi-talker multi-channel meeting-style datasets AMI~\cite{carletta2005ami}, AliMeeting (AliM)~\cite{yu2022m2met}, AISHELL-4 (ASH)~\cite{fu2021aishell}, and NOTSOFAR-1 (NSF) \cite{vinnikov2024notsofar}.
In all multi-channel experiments, four microphones are used, irrespective of the total number of microphones available in the respective datasets. Microphones are selected such that the spacing between microphones is maximized to ensure optimal capture of spatial information.
All models employ the WavLM Base+ from~\cite{han2025efficient_journal} as the single-channel FM, which is pruned and fine-tuned on the datasets used in this work, as well as additional data.
During the training process of the diarization system, this FM is also finetuned further with a smaller learning rate.
\end{sloppypar}

Tests are conducted on the above four datasets, plus on DiPCo \cite{dipco}. As DiPCo was not part of the training, it serves to assess the generalization capability of the systems to unseen datasets and microphone geometries. 
For the evaluation, we report the diarization error rate (DER), where no collar is used. The segment-level EEND outputs are grouped with oracle clustering to resolve permutation ambiguity between the segments.  To do so, the estimated local speaker activity is compared to the ground-truth activity to relate local speaker labels to ground-truth speaker labels. Using oracle clustering allows for concentrating on the impact of spatial features on the local EEND module, independent of the performance of the subsequent clustering module.

\section{Results}
\label{sec:results}

\subsection{Comparison of Multi-Channel Approaches}

\begin{table}
    \centering
    \caption{DER of the baseline and the different multi-channel extensions in \%. Results are shown for single-speaker regions (Single), overlapping speech (OV), and overall (Total).}%
    \label{tab:results_1}
    \setlength{\tabcolsep}{0.9pt}
    \sisetup{round-mode=places,round-precision=1}
    \begin{tabular}{lSSSSSSSS} %
    \toprule

\multirow{2}{*}{\textbf{Type}} 
& {\multirow{2}{*}{\textbf{AMI}} }
& {\multirow{2}{*}{\textbf{AliM}} }
& {\multirow{2}{*}{\textbf{ASH}}} 
& {\multirow{2}{*}{\textbf{NSF}} }
& {\multirow{2}{*}{\textbf{DiPCo}} } %
& \multicolumn{3}{c}{\textbf{Macro}} \\
\cmidrule(lr){7-9}
& & & & & &  \textbf{Single} & \textbf{OV} & \textbf{Total} \\
    \midrule
    Baseline\cite{han2025leveraging} & 13.20 & 12.66 & 9.25 & 14.49 & 31.8 & 12.2 & 23.9 & 16.28 \\  %
    \quad + Beamf. &  13.30 & 12.94 & 8.99 & 14.52 & 33.0 & 12.0 & 25.0 & 16.54  \\  %
    \midrule 
    MC-FM\cite{han2025spatial}& 13.0 & \bfseries 11.7 & 9.1 & 14.2 & 32.5 & 12.2 & \bfseries 23.5 & 16.1\\
    \midrule 
    Condition.~\cite{hscma} & \bfseries 12.2 & 11.79 & \bfseries 8.9 & \bfseries 13.4  & \bfseries 30.3 & \bfseries 11.0 & \bfseries 23.5 & \bfseries15.3\\  %
     \quad + Beamf. &  12.52 & 11.87 & 8.96 & 13.71 & 32.4 & \bfseries 11.0 & 25.0 & 15.90 \\ %
    \bottomrule
    \end{tabular}
\end{table}

\Cref{tab:results_1} shows the DER of the different approaches. While the multi-channel DiariZen (MC-FM) led to some improvement of DER compared to the baseline system, the best performance of \SI{15.3}{\percent} DER was achieved with the explicit multi-channel conditioning.
Therefore, the explicit use of spatial features as conditioning instead of an implicit use via multi-channel FM, is advantageous for the diarization task and achieves the best result of all considered systems.

Apparently, the cascade of BeamformIt and the single-channel FM does not lead to any improvement in DER, same when using the BeamformIt output instead of the reference channel in the conditioning approach.
A possible explanation is that while the beamformer exploits the spatial diversity of the target source and the distortions to improve the signal quality, its behavior in case of two or more active speakers may even counteract the diarization objective: in regions of speech overlap, any spatial cues helpful to identify that more than a single speaker is active, are lost at the beamformer output. 
This can be seen when comparing the DER for the baseline system calculated separately on single and overlapping speaker segments, where single speaker results improve by $0.2$ percentage points, but the DER increases by $1.1$ percentage points in overlap regions, resulting in a lower overall diarization performance. 
This explanation is corroborated when looking at the breakdown of the results per dataset: An improvement is only seen for AISHELL-4, which is the dataset with the smallest ratio of overlapped speech, whereas the larger amount of overlapping speech such as in AMI or AliMeeting incurs higher overall DER. %

Note that in \cite{cspb} the use of BeamformIt led to improvements and that BeamformIt with WavLM was among the top-performing systems. To understand these contrary findings and rule out possible errors on our side we repeated our test mimicking the training conditions of \cite{cspb}, where training was only done on AMI and with a vanilla WavLM. Under these conditions, we achieved similar results as in \cite{cspb}.

Furthermore, the systems are evaluated on the previously unseen dataset DiPCo. Here, the multi-channel DiariZen does not perform better than the baseline DiariZen with \SI{32.5}{\percent} and \SI{31.8}{\percent} DER, respectively.
The conditioned system achieves the  best performance on DiPCo with \SI{30.3}{\percent} DER.
Therefore, the conditioning system not only achieves the best results on datasets seen during training, but also shows generalization capabilities to unseen datasets and is able to generalize to unseen spatial characteristics and microphone array geometries, thus supporting the claim in \cite{hscma}.

\begin{table}
\setlength{\tabcolsep}{5pt}
\centering
\caption{Distribution of error frames across systems. Each entry denotes the percentage of all error frames in which the specified subset of systems (spatial, spectral, combined) produces
an error at the same time frame. Results are shown for single-speaker regions (Single), overlapping speech (OV), and overall (Total).}
\label{tab:error_frames_transposed}
\begin{tabular}{lSSSS|ccc}
\toprule
{\textbf{Systems}} 
& \textbf{Single} & \textbf{OV} & \textbf{Total} \\
\midrule

spatial \& spectral \& combined
& 34.18 & 57.30 & 46.99 \\

spatial \& spectral 
& 4.10 & 3.72 & 3.76 \\

spatial \& combined
& 3.54 & 4.28 & 3.63 \\

spectral \& combined
& 15.85 & 7.03 & 12.31 \\

spatial only 
& 29.72 & 21.01 & 23.71 \\

spectral only 
& 9.04 & 4.62 & 6.62 \\

combined only 
& 3.58 & 2.06 & 3.00 \\

\bottomrule
\end{tabular}
\end{table}

\subsection{Error Analysis}
\label{sec:error_analysis}

In the preceding experiments, the conditioning system that first computes both spectral and spatial features and then merges them via FiLM layers turned out to deliver the lowest DER. 
Luckily, this system is also the one that is best accessible to an analysis of the benefits of spectral, spatial, and jointly spectral and spatial information for diarization, because it has separate spectral and spatial feature extraction components, where each of them can drive the diarization alone. 
To this end, we study the conditioning system %
and refer to it as the \textit{combined} system, and to its components as the \textit{spatial} and \textit{spectral} systems. Note that the spectral system component corresponds to the single-channel DiariZen baseline and the spatial component corresponds to the auxiliary multi-channel network (i.e. the spatial diarization system from~\cite{hscma}).

In \cref{tab:error_frames_transposed} the distribution of error frames across systems is analyzed.
To investigate the degree of shared and component-specific errors, the error frames are partitioned based on which subset of systems produces an error at a given time frame. This allows us to distinguish between errors that only occur in one system component (\textit{spectral only}, \textit{spatial only}, \textit{combined only}), pairwise shared errors (\textit{spatial \& spectral}, \textit{spatial \& combined}, \textit{spectral \& combined}), and errors common to all systems (\textit{spatial \& spectral \& combined}).
All values are reported as percentages relative to the total number of frames containing at least one error.

This decomposition demonstrates that the combined model does not merely inherit the errors of the individual systems but is able to correct errors that occur in only one of the separate systems and can even correct some errors that occur in both systems at the same time. 
It corrects \SI{6.62}{\percent} of error frames that are unique to the spectral system (\textit{spectral only}), \SI{23.71}{\percent} that are unique to the spatial system (\textit{spatial only}) and \SI{3.76}{\percent} of errors that occur in both systems simultaneously (\textit{spatial \& spectral}).

However, the combined model does not fully eliminate the component-specific errors. A subset of errors originating from the spatial system persists in the combined model (\SI{3.63}{\percent}, \textit{spatial \& combined}), and similarly for the spectral system (\SI{12.31}{\percent}, \textit{spectral \& combined}).
Furthermore, the combined system introduces a small proportion of new errors that are not present in either individual model (\SI{3}{\percent}, \textit{combined only}).
Overall, the combined system corrects more errors of the individual systems than it introduces new errors and therefore improves the performance.
Since the majority of the persisting errors stems from the overlap with the spectral system (\SI{12.3}{\percent}, \textit{spectral \& combined}) compared to the spatial system (\SI{3.6}{\percent}, \textit{spatial \& combined}), the error patterns of the combined model appear to be more similar to those of the spectral system.
This observation is consistent with the stronger overall performance of the spectral system.

In overlap regions, the proportion of subsystem-specific errors is reduced, whereas a larger fraction of errors is shared across all systems (\SI{57.3}{\percent}), indicating more consistent failure modes under overlap conditions.
In contrast, single-speaker regions exhibit more diverse error patterns. 
The proportion of errors shared by all systems decreases to \SI{34.18}{\percent}, while subsystem-specific errors increase across all models, suggesting that errors occur more independently across systems.
Furthermore, the spectral system contributes a smaller share of the total errors (\SI{11.6}{\percent}, \textit{spectral only} + \textit{spectral \& combined}) in the overlap regions than the spatial system (\SI{25.3}{\percent}, \textit{spatial only} + \textit{spatial \& combined}).
This is somewhat surprising, given that spatial features are generally considered beneficial for handling overlapping speech.

\subsection{Concurrent Speaker Detection using WavLM}

The error analysis of the last section revealed a surprisingly strong performance of the spectral-only diarization system in regions of overlapping speech, where more than one speaker is active. It performed even better than one relying on spatial features. 
This is somewhat against expectations for two reasons: first, because spatial features are often believed to be particularly helpful in the case of speech overlap \cite{cordlandwehr25_interspeech,wang2022spatial,chen21t_interspeech,yoshioka18_interspeech,zelenak10_interspeech_spatial_overlap} 
and, second, since the spectral features are derived from WavLM, where WavLM had originally been trained with a masked prediction loss to suppress a competing speaker \cite{Chen2022_WavLM}.

To better understand this behavior, we further investigate the overlap-related information encoded in the WavLM representations using a Concurrent Speaker Detection (CSD) task as introduced in~\cite{sharon_csd}. This task directly models the presence of zero (class 0), one (class 1), or multiple (class 2) active speakers per time frame and thus isolates the overlap detection capability that is implicitly required in speaker diarization as the powerset loss of the  EEND module assumes a maximum of 2 concurrent speakers per frame.
Thus, CSD closely reflects the core frame-level decisions required for speaker diarization, i.e., voice activity detection and overlap detection.

To isolate the capabilities of the WavLM features, we employ a lightweight MLP head consisting of two linear layers with a ReLU activation in-between. The WavLM backbone is kept frozen in all CSD experiments. 
In addition to the pruned WavLM from~\cite{han2025efficient_journal} that was used in preceding experiments, we also evaluated the vanilla WavLM Base+ to see the impact the finetuning on a diariation objective of the pruned WavLM has on the CSD performance.
The experiment is also performed with the frozen spatial conformer auxiliary multi-channel network from~\cite{hscma} as feature extractor to compare against a learned multi-channel system.

The resulting F1-scores for the CSD task are shown in \cref{tab:csd} for the vanilla WavLM, the pruned WavLM, and the spatial conformer.
For reference, we compare against a multi-microphone transformer-based dedicated CSD system from~\cite{sharon_csd}.\footnote{Since in~\cite{sharon_csd} only the confusion matrices and class frequencies were published, the presented F1-scores have been calculated from those values.} 
Although not directly comparable, because the system from~\cite{sharon_csd} was trained only on AMI and AliMeeting, the results give an indication how powerful the single-channel spectral WavLM features are for the CSD task, compared to a multi-channel spatial approach.
The vanilla WavLM achieves a macro F1-score of $0.8$ on AMI and $0.8$ on AliMeeting, outperforming the multi-channel system with F1-scores of $0.7$ and $0.7$, respectively. 
In overlap, the multi-channel CSD achieves a macro F1-score of $0.6$, while the vanilla WavLM model reaches $0.7$ across all datasets.
Using the pruned WavLM model further improves performance to a macro F1-score of $0.8$ and an overlap F1-score of $0.8$, showing that the diarization objective further increases CSD performance.
The spatial conformer, which was also pre-trained on a diarization objective, performs similarly to vanilla WavLM.

Interestingly, the WavLM model achieves strong overlap detection performance despite using only single-channel data and despite being trained with a masked speech prediction objective, which learns to reconstruct masked speech from contextual information.
Such a training objective could be expected to bias the representations towards the dominant speaker in overlapping conditions, while suppressing the competing one.
This indicates that WavLM features, when aggregated across layers of its transformer blocks, nevertheless contain information relevant to overlapping speech.

\begin{table}[t]
\setlength{\tabcolsep}{2pt}
    \centering
    \caption{F1-scores on the CSD task for different feature extractors evaluated on AMI, AISHELL-4, AliMeeting, and NOTSOFAR-1.}
    \label{tab:csd}
    \begin{tabular}{lSSSSSS}
    \toprule
         
\multirow{2}{*}{\textbf{Systems}} 
& {\multirow{2}{*}{\textbf{AMI}}} 
& {\multirow{2}{*}{\textbf{AliM}} }
& {\multirow{2}{*}{\textbf{ASH}} }
& {\multirow{2}{*}{\textbf{NSF}} }
& \multicolumn{2}{c}{\textbf{Macro}} \\
& & & & & {\textbf{total}} &{ \textbf{OV}} \\
    \midrule
       Vanilla WavLM~\cite{Chen2022_WavLM} & 0.791 & 0.819 & 0.747 & 0.842 & \textbf{0.8} & 0.67675\\
       Pruned WavLM~\cite{han2025efficient_journal} & \textbf{0.9} & \textbf{0.9} & \textbf{0.8} & \textbf{0.9} & \textbf{0.8} & \textbf{0.8}\\
       \midrule
       \makecell[l]{Multi-microphone\\transformer-based~\cite{sharon_csd}} & 0.6639 & 0.7334  & {-} & {-} & 0.69865 & 0.5878 \\
    \midrule
        Spatial conformer~\cite{hscma} & 0.838 & 0.759 & \textbf{0.8} & 0.837 & \textbf{0.8} & 0.71675 \\ 
    \bottomrule
    \end{tabular}
\end{table}

\section{Conclusion}
\label{sec:conclusion}
We analyzed three approaches to combine multi-channel input with foundation-model-based diarization systems. Building upon the DiariZen baseline, our results show that a cascaded approach using a beamformer as a preprocessing step is
detrimental to the performance in overlap regions. In contrast, native multi-channel processing is able to yield consistent performance gains, while the best performance is achieved by conditioning the downstream network with explicit spatial cues. A  detailed error analysis of this conditioning showed that errors made by using either only spectral or spatial information can be overruled to some extent, while introducing a comparatively small number of new errors through the combination. It turned out that, surprisingly, the spectral model demonstrated stronger performance in overlap conditions compared to the spatial model. Overall, the comparison suggests that the combination of an explicit spatial processor with a single-channel foundation model is a competitive approach to diarization, which we aim to extend to other multi-channel speech processing tasks in future work.

\bibliographystyle{IEEEbib}
\balance
\newpage
\bibliography{refs}

@string{icassp = "Proc. IEEE ICASSP"}

@string{interspeech = "Proc. ISCA Interspeech"}

@string{ieee-taslp = "IEEE Trans. Audio, Speech, Lang. Process."}

@string{ieee-jstsp = "IEEE J. Sel. Top. Signal Process."}

@string{eusipco = "Proc. EUSIPCO"}

@string{hscma = "Proc. HSCMA"}

@string{slt = "Proc. IEEE SLT"}

@string{arxiv = "ArXiv"}

@string{eacl = "Proc. EACL"}

@inproceedings{bredin23_pyannote,
  title     = {{pyannote.audio 2.1 speaker diarization pipeline: principle, benchmark, and recipe}},
  author    = {Hervé Bredin},
  year      = {2023},
  booktitle = {{Interspeech 2023}},
  pages     = {1983--1987},
  doi       = {10.21437/Interspeech.2023-105},
  issn      = {2958-1796},
}

@INPROCEEDINGS{wang2022spatial,
  title={Spatial-aware speaker diarization for multi-channel multi-party meeting},
  author={Wang, Jie and Liu, Yuji and Wang, Binling and Zhi, Yiming and Li, Song and Xia, Shipeng and Zhang, Jiayang and Tong, Feng and Li, Lin and Hong, Qingyang},
  booktitle =interspeech,
  year={2022}
}

@INPROCEEDINGS{sharon_csd,
  author={Eliav, Amit and Gannot, Sharon},
  booktitle={2024 32nd European Signal Processing Conference (EUSIPCO)}, 
  title={Concurrent Speaker Detection: A Multi-Microphone Transformer-Based Approach}, 
  year={2024},
  volume={},
  number={},
  pages={897-901},
  doi={10.23919/EUSIPCO63174.2024.10715386}}

@inproceedings{gulati20_interspeech,
  title     = {Conformer: Convolution-augmented Transformer for Speech Recognition},
  author    = {Anmol Gulati and James Qin and Chung-Cheng Chiu and Niki Parmar and Yu Zhang and Jiahui Yu and Wei Han and Shibo Wang and Zhengdong Zhang and Yonghui Wu and Ruoming Pang},
  year      = {2020},
  booktitle = interspeech,
  pages     = {5036--5040},
  doi       = {10.21437/Interspeech.2020-3015},
  issn      = {2958-1796},
}

@inproceedings{cordlandwehr25_interspeech,
  title     = {{Spatio-Spectral Diarization of Meetings by Combining {TDOA}-based Segmentation and Speaker Embedding-based Clustering}},
  author    = {Tobias Cord-Landwehr and Tobias Gburrek and Marc Deegen and Reinhold Haeb-Umbach},
  year      = {2025},
  booktitle = interspeech,
  pages     = {5223--5227},
  doi       = {10.21437/Interspeech.2025-1663},
  issn      = {2958-1796},
}

@inproceedings{kinoshita21_interspeech,
  title     = {Advances in Integration of End-to-End Neural and Clustering-Based Diarization for Real Conversational Speech},
  author    = {Keisuke Kinoshita and Marc Delcroix and Naohiro Tawara},
  year      = {2021},
  booktitle = interspeech,
  pages     = {3565--3569},
  doi       = {10.21437/Interspeech.2021-1004},
  issn      = {2958-1796},
}

@inproceedings{luo2020end,
  title={End-to-end microphone permutation and number invariant multi-channel speech separation},
  author={Luo, Y. and Chen, Z. and Mesgarani, N. and others},
  booktitle=icassp,
  pages={6394--6398},
  year={2020}
}

@INPROCEEDINGS{yu2022m2met,
  author={Yu, Fan and Zhang, Shiliang and Fu, Yihui and Xie, Lei and Zheng, Siqi and Du, Zhihao and Huang, Weilong and Guo, Pengcheng and Yan, Zhijie and Ma, Bin and Xu, Xin and Bu, Hui},
  booktitle=icassp, 
  title={M2Met: The Icassp 2022 Multi-Channel Multi-Party Meeting Transcription Challenge}, 
  year={2022},
  volume={},
  number={},
  pages={6167-6171},
  doi={10.1109/ICASSP43922.2022.9746465}}

@inproceedings{vinnikov2024notsofar,
  title     = {{NOTSOFAR-1 Challenge: New Datasets, Baseline, and Tasks for Distant Meeting Transcription}},
  author    = {Alon Vinnikov and Amir Ivry and Aviv Hurvitz and Igor Abramovski and Sharon Koubi and Ilya Gurvich and Shai Peer and Xiong Xiao and Benjamin Martinez Elizalde and Naoyuki Kanda and Xiaofei Wang and Shalev Shaer and Stav Yagev and Yossi Asher and Sunit Sivasankaran and Yifan Gong and Min Tang and Huaming Wang and Eyal Krupka},
  year      = {2024},
  booktitle = interspeech,
  pages     = {5003--5007},
  doi       = {10.21437/Interspeech.2024-1788},
  issn      = {2958-1796},
}

@inproceedings{fu2021aishell,
  title={{AISHELL-4}: An open source dataset for speech enhancement, separation, recognition and speaker diarization in conference scenario},
  author={Fu, Y. and Cheng, L. and Lv, S. and others},
  booktitle=interspeech,
  pages={3665--3669},
  year={2021}
}

@inproceedings{carletta2005ami,
  title={The {AMI} meeting corpus: A pre-announcement},
  author={Carletta, J. and Ashby, S. and others},
  booktitle={Proc. MLMI},
  pages={28--39},
  year={2005}
}

@article{film,
title={FiLM: Visual Reasoning with a General Conditioning Layer}, volume={32}, url={https://ojs.aaai.org/index.php/AAAI/article/view/11671}, 
DOI={10.1609/aaai.v32i1.11671}, 
number={1}, 
journal={Proc. AAAI Conference on Artificial Intelligence},
author={Perez, Ethan and Strub, Florian and de Vries, Harm and Dumoulin, Vincent and Courville, Aaron},
year={2018}, 
}

@INPROCEEDINGS{kinoshita21_icassp,
  author={Kinoshita, Keisuke and Delcroix, Marc and Tawara, Naohiro},
  booktitle=icassp, 
  title={Integrating End-to-End Neural and Clustering-Based Diarization: Getting the Best of Both Worlds}, 
  year={2021},
  volume={},
  number={},
  pages={7198-7202},
  doi={10.1109/ICASSP39728.2021.9414333}}

@ARTICLE{Chen2022_WavLM,
  author={Chen, Sanyuan and Wang, Chengyi and Chen, Zhengyang and Wu, Yu and Liu, Shujie and Chen, Zhuo and Li, Jinyu and Kanda, Naoyuki and Yoshioka, Takuya and Xiao, Xiong and Wu, Jian and Zhou, Long and Ren, Shuo and Qian, Yanmin and Qian, Yao and Wu, Jian and Zeng, Michael and Yu, Xiangzhan and Wei, Furu},
  journal=ieee-jstsp, 
  title={WavLM: Large-Scale Self-Supervised Pre-Training for Full Stack Speech Processing}, 
  year={2022},
  volume={16},
  number={6},
  pages={1505-1518},
  doi={10.1109/JSTSP.2022.3188113}}

@inproceedings{han2025leveraging,
  title={Leveraging self-supervised learning for speaker diarization},
  author={Han, Jiangyu and Landini, Federico and Rohdin, Johan and Silnova, Anna and Diez, Mireia and Burget, Luk{\'a}{\v{s}}},
  booktitle=icassp,
  year={2025}
}

@article{Raj2021Doverlap,
  title={{DOVER-Lap}: A Method for Combining Overlap-aware Diarization Outputs},
  author={D.Raj and P.Garcia and Z.Huang and S.Watanabe and D.Povey and A.Stolcke and S.Khudanpur},
  journal=slt,
  year={2021}
}

@inproceedings{han2025spatial,
  title={Spatially Aware Self-Supervised Models for Multi-Channel Neural Speaker Diarization}, 
  author={Jiangyu Han and Ruoyu Wang and Yoshiki Masuyama and Marc Delcroix and Johan Rohdin and Jun Du and Lukas Burget},
  year={2025},
  eprint={arXiv preprint arXiv:2510.14551},
  booktitle={arXiv preprint},
  primaryClass={eess.AS},
  url={https://arxiv.org/abs/2510.14551}, 
}

@ARTICLE{han2025efficient_journal,
  author={Han, Jiangyu and Pálka, Petr and Delcroix, Marc and Landini, Federico and Rohdin, Johan and Černocký, Jan and Burget, Lukáš},
  journal=ieee-taslp, 
  title={Efficient and Robust Speaker Diarization via Structured Pruning of Self-Supervised Models}, 
  year={2026},
  volume={34},
  number={},
  pages={1903-1914},
}

@article{beamformit,
  author = {X. Anguera and C. Wooters and J. Hernando},
  title = {Acoustic beamforming for speaker diarization of meetings},
  journal = ieee-taslp,
  year = {2007},
  volume = {15},
  pages = {2011-2021},
  number = {7},
}

@inproceedings{cspb,
    title = "{CSPB}: Conversational Speech Processing Benchmark for Self-supervised Speech Models",
    author = "Huang, Zili  and
      Maciejewski, Matthew  and
      Garcia Perera, Leibny Paola  and
      Watanabe, Shinji  and
      Khudanpur, Sanjeev",
    booktitle = eacl,
    year = "2026",
    pages = "5878--5893",
}

@inproceedings{hscma,
  title={On the Role of Spatial Features in Foundation-Model-Based Speaker Diarization}, 
  author={Deegen, Marc and Gburrek, Tobias and Cord-Landwehr, Tobias and von Neumann, Thilo and Han, Jiangyu and Burget, Lukas and Haeb-Umbach, Reinhold},
  year={2026},
  eprint={arXiv preprint arXiv:2601.02231},
  booktitle={arXiv preprint},
  primaryClass={eess.AS},
  url={https://arxiv.org/abs/2601.02231}, 
}

@INPROCEEDINGS{unixenc,
  author={Huang, Zili and Shao, Yiwen and Zhang, Shi-Xiong and Yu, Dong},
  booktitle=icassp, 
  title={UniX-Encoder: A Universal X-Channel Speech Encoder for AD-HOC Microphone Array Speech Processing}, 
  year={2024},
  volume={},
  number={},
  pages={11991-11995},
  doi={10.1109/ICASSP48485.2024.10448072}}

@inproceedings{dipco,
  title     = {{DiPCo — Dinner Party Corpus}},
  author    = {Maarten Van Segbroeck and Ahmed Zaid and Ksenia Kutsenko and Cirenia Huerta and Tinh Nguyen and Xuewen Luo and Björn Hoffmeister and Jan Trmal and Maurizio Omologo and Roland Maas},
  year      = {2020},
  booktitle = interspeech,
  pages     = {434--436},
  doi       = {10.21437/Interspeech.2020-2800},
  issn      = {2958-1796},
}

@inproceedings{chen21t_interspeech,
  title     = {{Overlapped Speech Detection Based on Spectral and Spatial Feature Fusion}},
  author    = {Weiguang Chen and Van Tung Pham and Eng Siong Chng and Xionghu Zhong},
  year      = {2021},
  booktitle = interspeech,
  pages     = {4189--4193},
  doi       = {10.21437/Interspeech.2021-2138},
  issn      = {2958-1796},
}

@inproceedings{zelenak10_interspeech_spatial_overlap,
  title     = {{Overlap detection for speaker diarization by fusing spectral and spatial features}},
  author    = {Martin Zelenák and Carlos Segura and Javier Hernando},
  year      = {2010},
  booktitle = {{Interspeech 2010}},
  pages     = {2302--2305},
  doi       = {10.21437/Interspeech.2010-631},
  issn      = {2958-1796},
}

@inproceedings{yoshioka18_interspeech,
  title     = {{Recognizing Overlapped Speech in Meetings: A Multichannel Separation Approach Using Neural Networks}},
  author    = {Takuya Yoshioka and Hakan Erdogan and Zhuo Chen and Xiong Xiao and Fil Alleva},
  year      = {2018},
  booktitle = interspeech,
  pages     = {3038--3042},
  doi       = {10.21437/Interspeech.2018-2284},
  issn      = {2958-1796},
}

\end{document}